\documentclass[twocolumn]{aastex631}

\newcommand{\qdfit}{\texttt{q3dfit}}

\newcommand{\um}{$\mu$m}

\newcommand{\feii}{[Fe II]}
\newcommand{\feiifull}{[Fe II] 5.34$\mu$m}

\newcommand{\oiii}{[\textrm{O}~\textsc{iii}]}

\newcommand{\ha}{${\rm H\alpha}$}

\newcommand{\jwst}{JWST}
\newcommand{\gaia}{Gaia}
\newcommand{\spitzer}{Spitzer}
\newcommand{\hst}{HST}

\shorttitle{VODKA-JWST: MIRI Spectroscopy of SDSS~J0749+2255}
\shortauthors{Chen et al.}
\begin{document}

%\title{Mid-Infrared Observation of a 4 kpc dual quasar at $z=2.17$ with JWST/MIRI }
%\title{Varstrometry for Off-nucleus and Dual Sub-kpc AGN (VODKA): JWST/Mid-InfraRed Instrument Integral-Field-Unit Spectroscopy of a 3.8 Kpc Dual Quasar at $z=2.17$}
\title{VODKA-JWST: A 3.8 kpc dual quasar at cosmic noon in a powerful starburst galaxy with JWST/MIRI IFU}

\newcommand{\jhu}{Department of Physics and Astronomy, Bloomberg Center, Johns Hopkins University, Baltimore, MD 21218, USA}
\newcommand{\stsci}{Space Telescope Science Institute, 3700 San Martin Drive, Baltimore, MD 21218, USA}
\newcommand{\ias}{Institute for Advanced Study, Princeton University, Princeton, NJ 08544, USA}
\newcommand{\uiuc}{Department of Astronomy, University of Illinois at Urbana-Champaign, Urbana, IL 61801, USA}

\correspondingauthor{Yu-Ching Chen}
\email{ycchen@jhu.edu}

\author[0000-0002-9932-1298]{Yu-Ching Chen}
\affiliation{\jhu}

\author[0000-0001-7572-5231]{Yuzo Ishikawa}
\affiliation{\jhu}

\author[0000-0001-6100-6869]{Nadia L. Zakamska}
\affiliation{\jhu}

\author[0000-0003-0049-5210]{Xin Liu}
\affiliation{\uiuc}
\affiliation{National Center for Supercomputing Applications, University of Illinois at Urbana-Champaign, Urbana, IL 61801, USA}
\affiliation{Center for Artificial Intelligence Innovation, University of Illinois at Urbana-Champaign, 1205 West Clark Street, Urbana, IL 61801, USA}

\author[0000-0003-1659-7035]{Yue Shen}
\affiliation{\uiuc}

\author[0000-0003-4250-4437]{Hsiang-Chih Hwang}
\affiliation{School of Natural Sciences, Institute for Advanced Study, 1 Einstein Drive, Princeton, NJ 08540, USA}

\author[0000-0002-1608-7564]{David Rupke}
\affiliation{Department of Physics, Rhodes College, 2000 N. Parkway, Memphis, TN 38112, USA}

\author[0000-0002-0710-3729]{Andrey Vayner}
\affiliation{\jhu}

\author[0000-0001-7681-9213]{Arran C. Gross}
\affiliation{\uiuc}

\author[0000-0003-3762-7344]{Weizhe Liu}
\affiliation{Department of Astronomy, Steward Observatory, University of Arizona, Tucson, AZ 85719, USA}

\author[0000-0003-2212-6045]{Dominika Wylezalek}
\affiliation{Zentrum für Astronomie der Universität Heidelberg, Astronomisches Rechen-Institut, Mönchhofstr 12-14, D-69120 Heidelberg, Germany}

\author[0000-0002-3158-6820]{Sylvain Veilleux}
\affiliation{Department of Astronomy and Joint Space-Science Institute, University of Maryland, College Park, MD 20742, USA}

\author[0000-0002-6948-1485]{Caroline Bertemes}
\affiliation{Zentrum für Astronomie der Universität Heidelberg, Astronomisches Rechen-Institut, Mönchhofstr 12-14, D-69120 Heidelberg, Germany}

\author[0009-0003-5128-2159]{Nadiia Diachenko}
\affiliation{\jhu}

\author[0000-0002-4419-8325]{Swetha Sankar}
\affiliation{\jhu}

%% Note that the \and command from previous versions of AASTeX is now
%% depreciated in this version as it is no longer necessary. AASTeX 
%% automatically takes care of all commas and "and"s between authors names.

%% AASTeX 6.31 has the new \collaboration and \nocollaboration commands to
%% provide the collaboration status of a group of authors. These commands 
%% can be used either before or after the list of corresponding authors. The
%% argument for \collaboration is the collaboration identifier. Authors are
%% encouraged to surround collaboration identifiers with ()s. The 
%% \nocollaboration command takes no argument and exists to indicate that
%% the nearby authors are not part of surrounding collaborations.

%% Mark off the abstract in the ``abstract'' environment. 
\begin{abstract}

%Studying the interaction between kilo-parsec (kpc) dual quasars and their host galaxies via spatially resolved mid-infrared (MIR) spectra was challenging before the arrival of JWST.
%Prior to \jwst, studying the interaction between supermassive black holes and their host galaxies through spatially resolved Mid-Infrared (MIR) spectra in kilo-parsec (kpc) dual quasars at high redshifts presented significant challenges. 

Dual quasars, two active supermassive black holes at galactic scales, represent crucial objects for studying the impact of galaxy mergers and quasar activity on the star formation rate (SFR) within their host galaxies, particularly at cosmic noon when SFR peaks.
We present \jwst/MIRI mid-infrared integral field spectroscopy of J074922.96+225511.7, a dual quasar with a projected separation of 3.8 kilo-parsec at a redshift $z$ of 2.17.  We detect spatially extended \feiifull\ and polycyclic aromatic hydrocarbon (PAH) 3.3\um\ emissions from the star formation activity in its host galaxy. 
%The bolometric luminosity is dominated by the quasar based on the spectral energy distribution. 
We derive the SFR of 10$^{3.0\pm0.2}$ M$_{\odot}$ yr$^{-1}$ using PAH 3.3\um, which is five times higher than that derived from the cutoff luminosity of the infrared luminosity function for galaxies at $z\sim2$.
While the SFR of J0749+2255 agrees with that of star-forming galaxies of comparable stellar mass at the same redshifts, its molecular gas content falls short of expectations based on the molecular Kennicutt-Schmidt law. This discrepancy may result from molecular gas depletion due to the longer elevated stage of star formation, even after the molecular gas reservoir is depleted. We do not observe any quasar-driven outflow that impacts PAH and \feii\ in the host galaxy based on the spatially resolved maps. From the expected flux in PAH-based star formation, the \feii\ line likely originates from the star-forming regions in the host galaxy. Our study highlights the stardust nature of J0749+2255, indicating a potential connection between the dual quasar phase and intense star formation activities.
%the capability of \jwst\ to conduct spatially resolved MIR observations for kpc dual quasars at $z\sim2$, enabling an examination of SFR, shock diagnostics, and molecular gas in their host galaxies.

\end{abstract}

%% Keywords should appear after the \end{abstract} command. 
%% The AAS Journals now uses Unified Astronomy Thesaurus concepts:
%% https://astrothesaurus.org
%% You will be asked to selected these concepts during the submission process
%% but this old "keyword" functionality is maintained in case authors want
%% to include these concepts in their preprints.
\keywords{Infrared astronomy(786) -- Double quasars(406) --- Active galactic nuclei(16) --- Galaxy mergers(608) --- Polycyclic aromatic hydrocarbons(1280) --- Star formation(1569)}

%% From the front matter, we move on to the body of the paper.
%% Sections are demarcated by \section and \subsection, respectively.
%% Observe the use of the LaTeX \label
%% command after the \subsection to give a symbolic KEY to the
%% subsection for cross-referencing in a \ref command.
%% You can use LaTeX's \ref and \label commands to keep track of
%% cross-references to sections, equations, tables, and figures.
%% That way, if you change the order of any elements, LaTeX will
%% automatically renumber them.
%%
%% We recommend that authors also use the natbib \citep
%% and \citet commands to identify citations.  The citations are
%% tied to the reference list via symbolic KEYs. The KEY corresponds
%% to the KEY in the \bibitem in the reference list below. 

\section{Introduction} \label{sec:intro}

Most nearby galaxies host supermassive black holes (SMBHs) at their centers \citep{Magorrian1998,KormendyHo2013}, and hierarchical structure formation and galaxy mergers are anticipated in the cold dark matter universe \citep{White1991,Navarro1996,Cole2000}. Following galaxy mergers, SMBHs sink into the center of the merged galaxy and undergo evolution through dynamical friction \citep{Yu2002,ChenYF2020}. Throughout this process, accretion via gas inflows could lead to significant electromagnetic waves from SMBHs, transforming them into active galactic nuclei or quasars \citep{hopkins08}. When both SMBHs are simultaneously active, emitting electromagnetic radiation, the system becomes a dual quasar. 

Identifying dual quasars on galactic scales, particularly at high redshifts ($z>0.5$), is challenging due to angular resolution limitations \citep[Figure 1 in][]{ChenYC2022}. Beyond $z>0.5$, the separation of two quasars within $<1\arcsec$ (corresponding to $\lesssim$8 kilo-parsec (kpc) at $z=1$) is below the typical resolving capability of ground-based optical/infrared telescopes. Conducting a systematic search for kpc dual quasars at $z>0.5$ using wide-field ground-based surveys is highly challenging, resulting in only a handful of such discoveries \citep{Inada12,Junkkarinen01,Schechter17,Lemon18,More16,Tang21}. Recently, the space-based \gaia\ survey has opened up new possibilities for finding kpc dual quasars, leading to the discovery of dozens of candidates \citep{HwangShen2020,ChenYC2022,Mannucci2022}. While some candidates proved to be star-quasar superpositions \citep{ChenYC2022} or single quasars gravitationally lensed into multiple images \citep{Gross2023,Li2023}, follow-up observations using various facilities have confirmed the dual quasar nature for a few of them \citep{Ciurlo2023,ChenYC2023a}.

Thanks to the high angular resolution (FWHM ranges from 0\farcs2 at 5.6 \um\ to 0\farcs8 at 25.5 \um) and sensitivity of the \jwst, the spatially resolved mid-infrared (MIR) observations of quasar host galaxies has become possible \citep{Rupke2023}. Various MIR emission features, including those from the polycyclic aromatic hydrocarbons (PAHs), molecular hydrogen and ionized gas, are diagnostic of gas physical conditions and excitation mechanisms in the host galaxies \citep{Shi2007,Armus2006} and therefore afford the first opportunity to investigate the interplay between quasar fueling, quasar feedback and star formation in its host. 

To explore kpc dual quasars in the MIR wavelengths for the first time, we perform \jwst\ Mid-InfraRed Instrument \citep[MIRI;][]{Rieke2015} Integral-Field-Unit (IFU) observations of a dual quasar, SDSSJ074922.96+225511.7 (hereinafter J0749+2255). 
%The study using NIRSpec data is presented in the other paper (Ishikawa et al. in prep.). 
This paper focuses on the analysis of MIRI data and on the study of detected emission features (e.g., \feiifull\ and PAH 3.3\um). We detail the observations, data reduction and analysis in Section \ref{sec:data}. 
We present the main results, including continuum maps, one-dimensional spectra, and dynamics maps of \feii\ and PAH emissions in Section \ref{sec:results}. The study then delves into investigating the star formation rate using PAH 3.3\um\ emission and whether \feii\ solely originates from star formation in Section \ref{sec:discussion}. We summarize our findings in Section \ref{sec:conclusion}.

\section{Target Selection, Observations, Data reduction and analysis} \label{sec:data}

\subsection{Our target: J0749+2255}

J0749+2255 is the first kpc dual quasar at $z=2.17$ with an identified host galaxy \citep{ChenYC2023a}. It was initially discovered using its variability-induced astrometric noise \citep{Shen2021,ChenYC2022}. The selecting technique, Varstrometry for Off-nucleus and Dual Sub-kiloparsec Active Galactic Nuclei (VODKA), utilizes the varibaiblity of quasars and the high astrometric accuracy of Gaia to find unresolved dual/lensed/off-nucleus quasars \citep{ShenHwang2019,HwangShen2020}. J0749+2255 was later confirmed as a dual quasar through multi-wavelength observations \citep{ChenYC2023a}. The spectra of both quasars in J0749+2255 are remarkably similar, and the Eddington ratios appear at least 0.1 \citep{ChenYC2023a}. J0749+2255 shows faint tidal tails in the deep HST near-IR images, as an indicator of a merging system \citep{ChenYC2023a}. However, recent JWST NIRSpec IFU observations reveal a giant rotating disk perpendicular to the direction of two nuclei, rather than a disturbed system with irregular merger morphology \citep{Ishikawa2024}.
Moreover, the \oiii\ emission appears faint, and there is no indication of robust galactic winds in the ionized gas \citep{Ishikawa2024}. 
All those findings make the formation of J0749+2255 and its relation with the host galaxy an intriguing topic.

\subsection{JWST MIRI/MRS data reduction}
J0749+2255 was observed with the Medium Resolution Spectrometer (MRS) mode of the MIRI instrument on \jwst\ on November 21-22, 2022 UT \citep{Argyriou2023}. The data were acquired in the long grating mode, which covers the wavelength range of 6.53--7.65\um, 10.02--11.70\um, 15.41--17.98\um, and 24.19--27.90\um\ in four channels. The observations utilized a 4-point dither pattern to enhance sampling and achieve improved spatial resolution in the final drizzled data cube \citep{Law2023}. Dedicated background exposures were simultaneously taken in the nearby field from the target position for background subtraction. All the JWST data used in this paper can be found in MAST: \dataset[10.17909/hh7a-c798]{http://dx.doi.org/10.17909/hh7a-c798}.

We process the data using \jwst\ calibration pipeline version 1.11.4 \citep{Bushouse2023.1.11.4}.
In the first stage, the \texttt{Detector1} pipeline applies detector-level corrections to all uncalibrated science and background exposures, converting raw images into corrected count rate files.
These count rate files are then processed with the \texttt{Spec2} pipeline, which performs various instrument-specific calibrations, including wavelength calibration, flat-field correction, flux calibration, fringing removal, and other calibrations \citep{Labiano2021}. We apply image-by-image background subtraction using background exposures taken simultaneously to account for flat-field correction and pixel-by-pixel variation in the detectors, rather than the default 1-dimensional (1D) master background subtraction. The \texttt{Spec2} pipeline converts count rate files into fully calibrated individual exposures.
In the \texttt{Spec3} pipeline, calibrated individual exposures across different wavelength bands and channels are combined into the final 3-dimensional drizzled data cubes.
Following the standard pipeline procedure, we eliminate faint strip patterns using procedures outlined in \citet{Spilker2023}. The stripe patterns were modeled using 2-dimensional (2D) background templates derived from the moving average of the data cube along the wavelength direction, with regions near the dual quasar excluded when creating the stripe templates.

\subsection{q3dfit fitting} \label{sec:q3dfit}
Revealing the faint emission from the host galaxy on top of the strong quasar emission might requires specialized treatment, involving dedicated Point Spread Function (PSF) subtraction and spectral decomposition techniques. In the analysis of J0749+2255, we use the cutting-edge software, \qdfit\  \citep{Rupke2014,Rupke2021}, available for \jwst\ through the Early-Release Science program ERS-01335 `Q3D'. \qdfit\ is capable of decomposing the emission of the quasar from that of the host galaxy and conducting spectral fitting for \jwst\ IFU data cubes. This tool has been utilized to study gas kinematics in quasar host galaxies in IFU data obtained with the \jwst\ \citep{Wylezalek2022,Vayner2023,Rupke2023} and from the ground \citep{Rupke2017}. 
Though \qdfit\ possesses the capability for PSF subtraction, we opt not to utilize it in this case. 
This decision is based on the observation that neither the PAH 3.3\um\ emission nor the \feiifull\ emission are dominated by the quasars, in that there is not a strong PSF component associated with the nuclei that would swamp the emission from the host.

To separate PAH 3.3\um\ features from the dust continuum, we 
fit the spectrum using a linear combination of a modified black-body function, representing the warm dust emission from the quasar's dusty torus, and the PAH template from  \citet{Lai2020}. This fitting is applied to the spectra in Channel 2 (10--11\um) at each spaxel. 
\qdfit uses a linear combination of pre-defined PAH templates to measure the contribution of PAHs to the MIR spectra, with only one fitting parameter, the amplitude, for each template. 
In practice, at wavelength $<$5\um\ (the wavelength range not covered by Spitzer observations), only one template from AKARI is available, which fits the 3.3 \um\ PAH feature with a linear combination of three Drude profiles \citep{Lai2020}.
The Drude profile is a theoretical profile for a classical damped harmonic oscillator, which is suitable for modeling PAH emission features \citep{Smith2007}. 
%However, to enhance the signal-to-noise ratio (S/N), we use the PAH template instead of a Drude model. 
The flux of the PAH 3.3\um\ feature is obtained by integrating the flux density of the PAH template between 3.2\um\ and 3.35\um. For the \feiifull\ line, we perform a local fit within a narrow wavelength range (16.65–17.2\um) in Channel 3. The continuum is modeled with a third-degree polynomial, and the \feii\ line is modeled with a single Gaussian function. 
%We remove the output result at the spaxel when the S/N ratio of the line intensity is less than 2, to avoid fitting to noise. 

\section{Results} \label{sec:results}

\subsection{MIR Continuum maps and spectra}

\begin{figure}
    \centering
    \includegraphics[width=0.45\columnwidth]{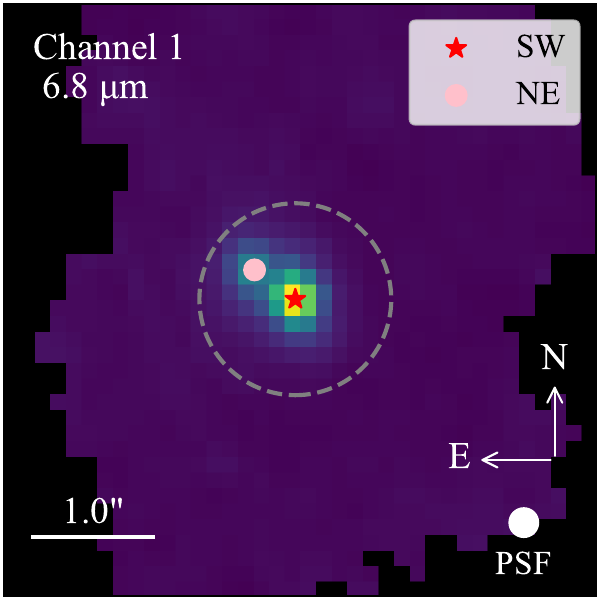}
    \includegraphics[width=0.45\columnwidth]{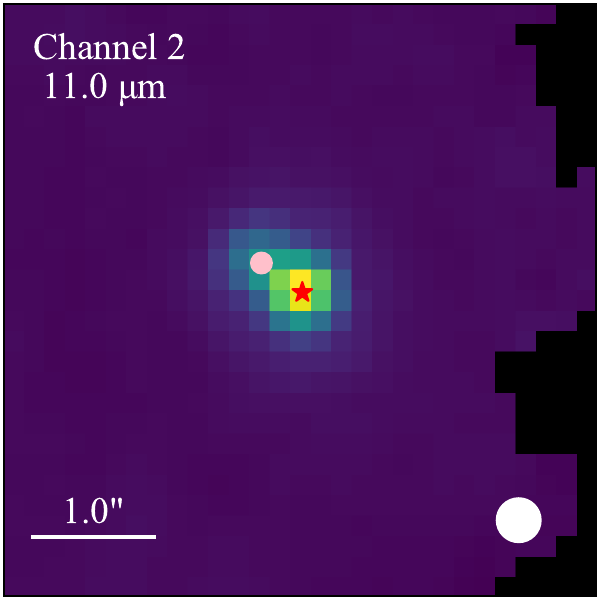}
    \includegraphics[width=0.45\columnwidth]{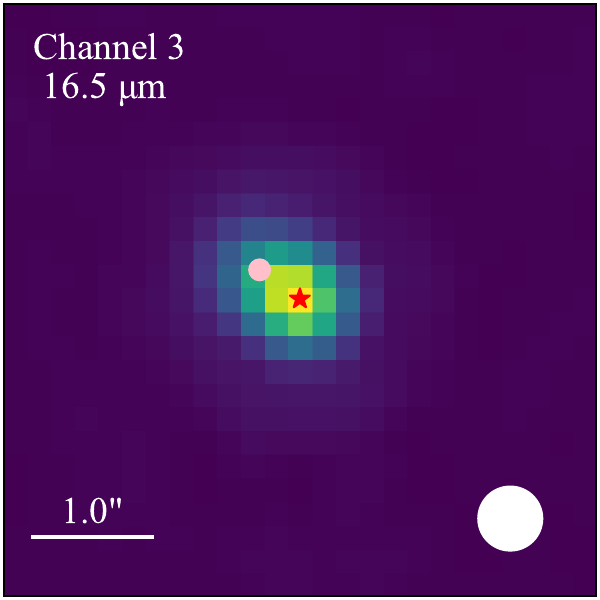}
    \includegraphics[width=0.45\columnwidth]{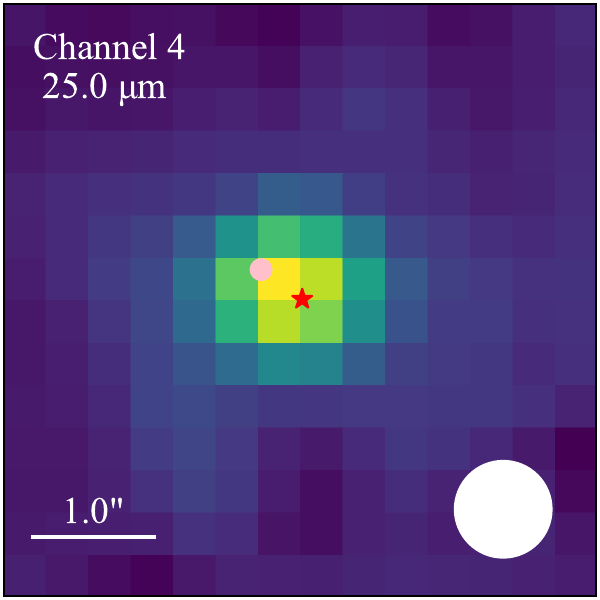}
    \caption{MIR Continuum images of J0749+2255 in the four channels. The color indicates the flux in linear scale. The positions of the quasars are indicated with pink circles and red stars. Two quasars are already marginally resolved in Channel 1. The black areas represent regions outside the MIRI Field of Views (FOVs). White filled circles represent the FWHM of the PSFs. A grey dashed circle denotes the aperture size utilized for extracting the spectra in \autoref{fig:spectra}.}
    \label{fig:cont_images}
\end{figure}

\begin{figure}
    \centering
    \includegraphics[width=0.9\columnwidth]{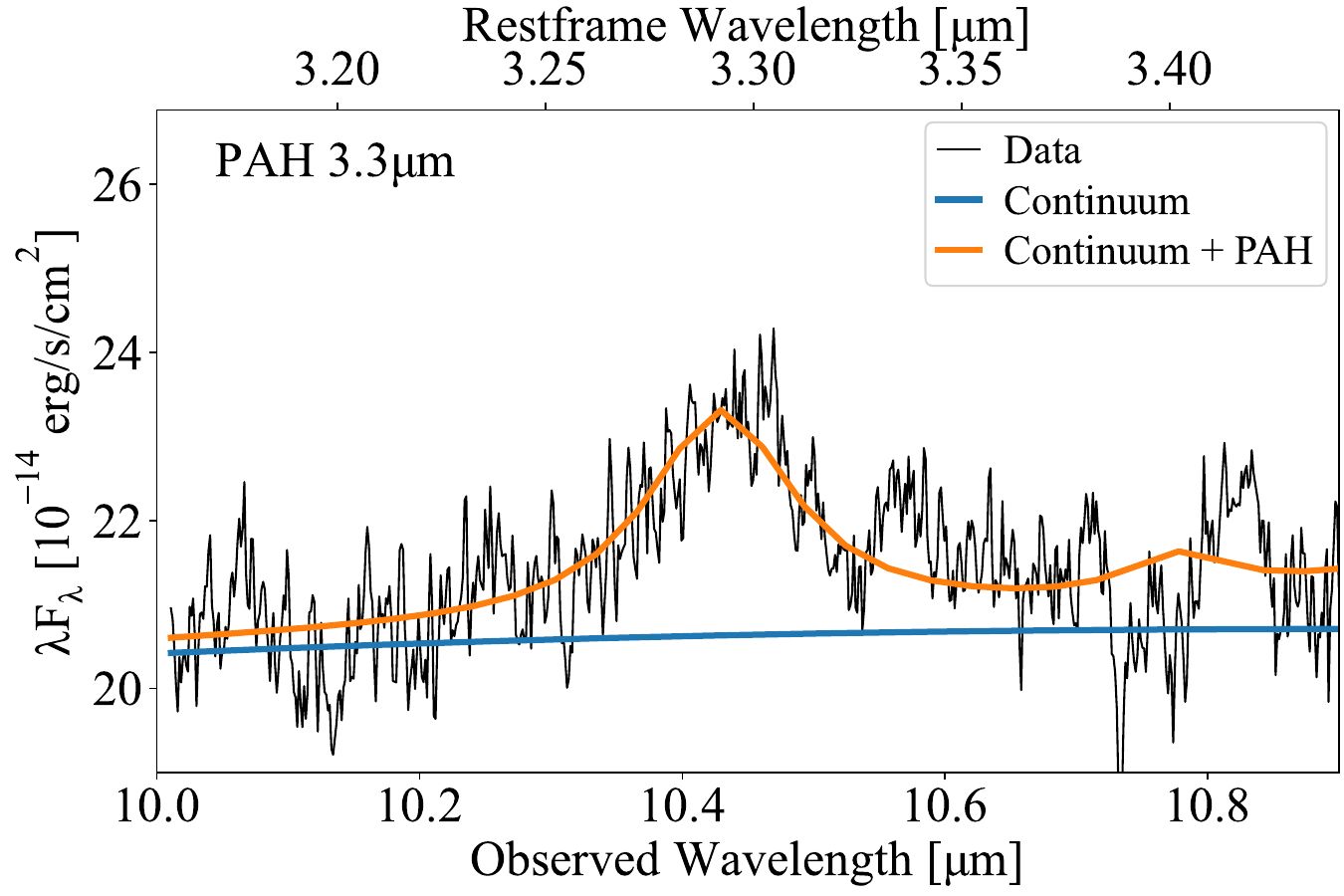}
    \includegraphics[width=0.9\columnwidth]{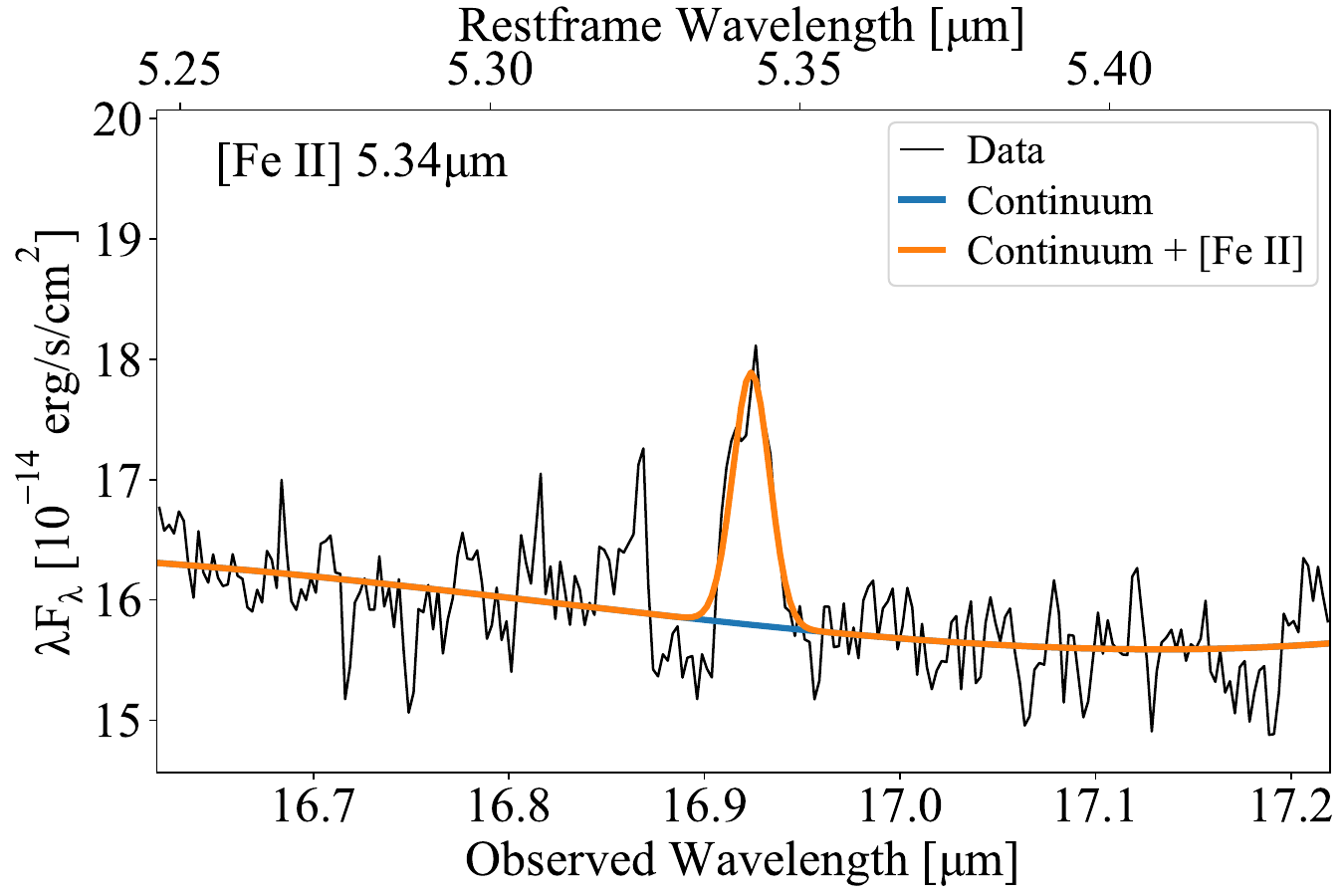}
    \caption{1D extracted MIR spectra of J0749+2255 centered at the PAH 3.3\um\ feature (Top) and the \feiifull\ line (Bottom). The spectra are for the entire system enclosed by the 0\farcs8 extraction aperture. The fitted models including a continuum, a PAH template, and/or a Gaussian component are also plotted.
    %Prominent features and emission lines are marked with red vertical lines. Unidentified features with unknown origins are marked with blue vertical lines. 
    }
    \label{fig:spectra}
\end{figure}

\autoref{fig:cont_images} shows MIR continuum images of J0749+2255 obtained from MIRI. 
Continuum maps were generated by averaging spectra within emission-line-free wavelength ranges. At the shortest wavelength of 7 \um, with an angular resolution of $\sim$0\farcs3 and a quasar separation of 0\farcs46, the dual quasar is marginally resolved in Channel 1. At longer wavelengths, the two nuclei appear blended into an elongated source due to lower spatial resolution. We use 2D Gaussian functions to fit the continuum images, with fixed positions and full-width-half-maxima (FWHM). The positions of the quasars were determined using the Channel 1 map because of its better spatial resolution, and the FWHM values were derived from diffraction-limited PSFs \citep{Rigby2023}. \autoref{tab:quasars} lists the fitted luminosities of two nuclei and their flux ratios.

\begin{deluxetable}{cccc}
 \tablecaption{Continuum luminosities of two nuclei and flux ratios at each channel. The errors are 1-$\sigma$ statistical uncertainties.
 \label{tab:quasars}}
 \tablehead{\colhead{Wavelength} & \colhead{log($\lambda$L$_{\lambda,\rm SW}$)} & \colhead{log($\lambda$L$_{\lambda,\rm NE}$)} & \colhead{Flux Ratio} \\
 \colhead{(\um)} & \colhead{(ergs/s)} & \colhead{(ergs/s)} & \colhead{} }
 \startdata
 6.8 & 45.44$\pm$0.04 & 45.11$\pm$0.08 & 2.1$\pm$0.4  \\
 11.0 & 45.55$\pm$0.03 & 45.26$\pm$0.06 & 1.9$\pm$0.3  \\
 16.5 & 45.52$\pm$0.03 & 45.22$\pm$0.06 & 2.0$\pm$0.3  \\
 25.0 & 45.58$\pm$0.10 & 45.37$\pm$0.17 & 1.6$\pm$0.7   
 \enddata
 %\tablecomments{}
\end{deluxetable}

Subsequently, we extract the 1D spectra for each channel using a circular region with a diameter of  0\farcs8 centered at the southwestern nuclei to identify the possible emission lines. We observe strong PAH features at 3.3\um\ and a significant \feiifull\ line. \autoref{fig:spectra} displays the 1D spectra of J0749+2255 centered at the PAH 3.3\um\ features and the \feiifull\ line. We also plot the fitted models on top of the data. Besides the PAH 3.3\um\ features and the \feiifull\ line, a faint PAH feature at 5.24\um\ and a faint rotational line of molecular hydrogen, H$_2$ 0–0 S(7), are likely detected. However, the fluxes of PAH 5.24\um\ and H$_2$ 0–0 S(7) are only slightly above the noise level, so we do not conduct further analysis for them.
%We also discover a few unidentified features between 3.5\um\ and 3.7\um\, for which we cannot match any known absorption lines observed in ultraluminous infrared galaxies (ULIRGs) or quasars. The spatial under-sampling of JWST IFU cube could produce wiggles in the single-spaxel spectrum, known as ``resampling noise" \citep{Smith2007resample,Law2023}. To understand if the detected features are real, we adjust the aperture size and location. Those features are persistent with different aperture sizes and disappear if far from quasars. Those unidentified features are unlikely artifacts due to resampling noise or stripes, but spectral leaks or other unknown aritifacts are still possible. 
In the subsequent sections, we focus on PAH 3.3\um\ and \feiifull\, exploring their morphology, line intensity, and dynamics.

\subsection{PAH 3.3$\mu$m map}

\begin{figure}
    \centering
    \includegraphics[width=0.6\columnwidth]{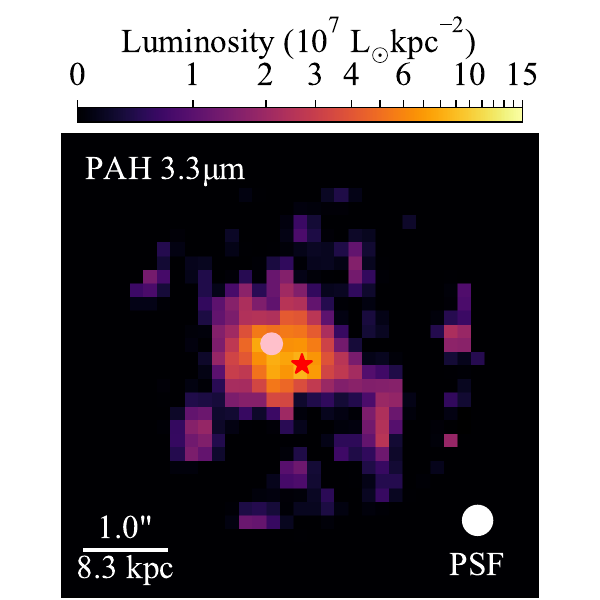}
    \caption{Integrated PAH 3.3$\mu$m luminosity map of J0749+2255 in units of 10$^7$ L$_\odot$ kpc$^{-2}$. North is up, and east is to the left. Spaxels close to the edges of IFU are masked to avoid the artifact features.}
    \label{fig:pah}
\end{figure}

We separate the PAH 3.3\um\ features from the dust continuum in Channel 2, as detailed in Section \ref{sec:q3dfit}. \autoref{fig:pah} shows the resulting PAH luminosity map. The PAH emission extends spatially out to approximately 10 kpc, particularly around the NE nucleus. By summing the luminosity within a radius of 1\farcs5 (equivalent to 12 kpc) at the center of the two nuclei, we find that the total luminosity of the PAH3.3\um\ feature is $L_{\rm PAH 3.3}$=10$^{9.8\pm0.1}$ L$_{\odot}$. The error represents 1-$\sigma$ systematic uncertainty based on fitting residuals, absolute flux calibration, and the accuracy of templates.

\subsection{[Fe II] line map}

\begin{figure}
    \centering
    \includegraphics[width=\columnwidth]{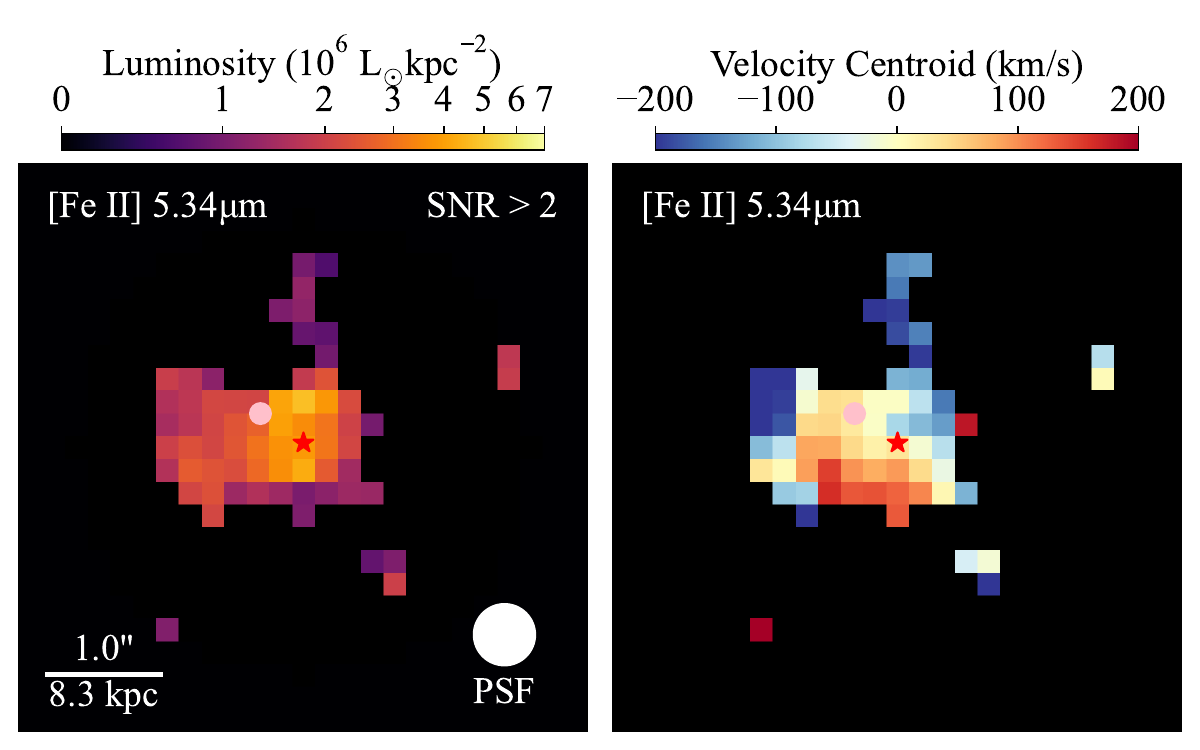}
    \caption{Left: Line-integrated \feiifull\ luminosity map of J0749+2255 in units of 10$^6$ L$_\odot$ kpc$^{-2}$. Right: Radial velocity centroid map in units of km s$^{-1}$. Spaxels close to the edges of IFU are masked to avoid the artifact features. Spaxels with \feii\ signal-to-noise ratio (S/N)$<$2 are also masked. North is up, and east is to the left.}
    \label{fig:feii}
\end{figure}
\autoref{fig:feii} shows \feiifull\ luminosity and velocity centroid maps. After subtracting the continuum, we find that the \feii\ emission exhibits an extended structure with a diameter of approximately 10 kpc. The \feii\ line is spectrally unresolved, with a typical fitted FWHM of  $\sim$150 km/s, close to the spectral resolution of MIRI at 17\um\ \citep{Labiano2021}. The velocity centroid map reveals an extended structure which is blueshifted toward the northwest and redshifted toward the southeast, with the velocity gradient perpendicular to the direction of the two nuclei. The radial velocities of the blueshifted and redshifted components are approximately 100-200 km/s. A similar kinematic pattern is also observed in the H$\alpha$ map which is interpreted by \citet{Ishikawa2024} as due to galaxy rotation. Using a 1\farcs5 radius aperture centered at the middle of two nuclei, we obtain the total luminosity of the \feii\ line $L_{\rm \feii}$=10$^{8.72\pm0.05}$ L$_{\odot}$. The error represents 1-$\sigma$ systematic uncertainty based on fitting residuals and absolute flux calibration.

\section{Discussion} \label{sec:discussion}

\subsection{Spectral Energy Distribution} \label{sec:dis:SED}

\begin{figure}
    \centering
    \includegraphics[width=\columnwidth]{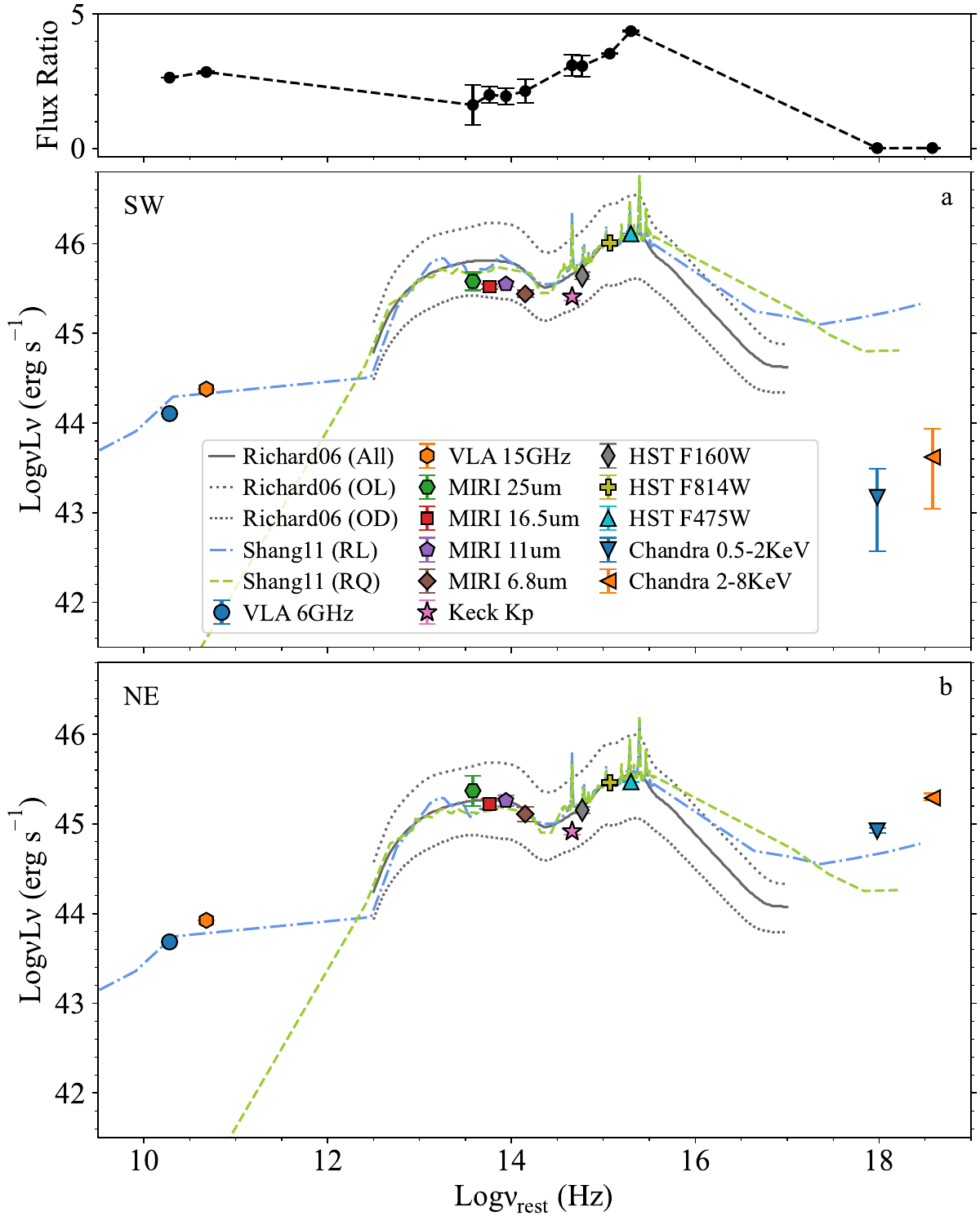}
    \caption{Top: SW/NE Flux ratios between two nuclei. Middle and Bottom: SEDs of SW and NE nuclei in J0749+2255. MIRI data points, along with multi-wavelength observations from \citet{ChenYC2023a}, are plotted. 
    The grey curves represent the SEDs of optically selected SDSS quasars \citep{richards06}, with solid lines showing mean values and dotted lines indicating optically luminous and dim sub-populations. Additional SEDs are from \citet{Shang2011}, covering both radio-loud and radio-quiet populations. The comparison templates are normalized to match the respective luminosities in the F814W filter of \hst\ observations.}
    \label{fig:SED}
\end{figure}

Leveraging the high angular resolution of the \jwst, we successfully separate the MIR continuum emissions of two nuclei. By incorporating values from \autoref{tab:quasars} and multi-wavelength observations from \citet{ChenYC2023a}, we present the spectral energy distributions (SEDs) of both nuclei in J0749+2255 in \autoref{fig:SED}. 
Both SEDs are dominant by emission from quasar and consistent with unobscured quasar templates \citep{richards06,Shang2011}, showing bumps in ultraviolet and MIR wavelengths. Previous radio observations also revealed that two nuclei both have very high radio luminosity of $\nu$L$_{\nu}\sim10^{43-44}$ erg s$^{-1}$ \citep{ChenYC2023a}. Combined with optical luminosity, the radio loudnesses $R_{\rm 6 cm/2500 A}$, defined as the flux ratio at the rest-frame 6 cm and that at 2500\AA, are more than 600. Our SEDs highlight the quasar-dominant and radio-loud nature of both nuclei.

The SW/NE flux ratios in J0749+2255 consistently decrease from 4.4 to 1.6 from rest-frame UV to MIR wavelengths. Then, the ratios rises to 2.8 in radio wavelengths. We already know from HST near-IR and Chandra X-ray observations that our target is most consistent with being a dual quasar rather than a lensed quasar \citep{ChenYC2023a}. Barring dramatic variability, any differences between the SEDs of the two quasars indicate that they comprise a physical quasar pair rather than a lensed quasar. The X-ray flux ratios are significantly lower ($\ll $1), further diminishing the likelihood of a lensed quasar scenario. Wavelength-dependent geometry or extinction in a lensed galaxy could introduce chromatic effects \citep{Barnacka2014,Pooley2007}, but the observed variations from X-ray to radio wavelengths make the lensed quasar scenario very unlikely.

\subsection{Star formation rate based on 3.3$\mu m$ PAH} \label{sec:dis:SFR}

PAH emission serves as an effective calorimeter for star formation rate (SFR; \citealt{Peeters2004,ForsterSchreiber2004}), even in quasar host galaxies where finding SFR measures unaffected by quasar radiation is very difficult \citep{Zakamska2016}. Ultraviolet photons from hot stars heat up PAH molecules, causing them to re-emit radiation in specific mid-infrared bands through vibration \citep{Puget1989}. 
Measures of SFR typically rely on stronger PAH features at longer wavelengths (e.g., 6.2, 7.7, 11.3 \um) for both quasar hosts \citep{Shi2007,Zakamska2016}  and star-forming galaxies \citep{Shipley2016,Xie2019}. Because of the sensitivity and wavelength coverage limitations of the \spitzer\ telescope, studies of SFR indicators using 3.3\um\ PAH remain scarce, especially for quasar hosts \citep{Kim2012,Lai2020}. We use the calibration of SFR for the 3.3 \um\ PAH from equation 1 in \citet{Lai2020}, 
\begin{equation}
    {\rm log}(\frac{\rm SFR}{\rm M_\odot yr^{-1}}) = -(6.80\pm0.18) + {\rm log}(\frac{L_{\rm PAH 3.3}}{L_{\odot}}),
\end{equation}
to compute the PAH-based SFR. The estimated total SFR for J0749+2255 is $10^{3.0\pm0.2}$ M$_{\odot}$ yr$^{-1}$. The 1-$\sigma$ error is dominant by the uncertainty in the conversion between PAH and SFR \citep{Lai2020}. 
The SFR of J0749+2255 is very high, $\sim$5 times higher than that derived from the cutoff luminosity $L_{\rm knee}$ of the infrared luminosity function for galaxies at $z\sim2$ \citep{Magnelli2013}. 

The SFR based on the 3.3\um\ PAH could be underestimated because small PAHs, traced by the PAH feature at shorter wavelengths as compared to the canonical 6.2\um\ and 7.7\um\ diagnostics, may be destroyed by the strong radiation from quasars \citep{Diamond-Stanic2010,Wu2010}. Recent JWST/MIRI IFU observation of a $z=4.22$ lensed sub-millimeter galaxy reveals a spatial mismatch between PAH 3.3\um\ and far-IR emission, demonstrating that PAH 3.3\um\ might not be a good direct indicator of the star formation rate of high-redshift galaxies \citep{Spilker2023}. However, some studies found no decline in the 3.3\um\ PAH intensity relative to the total PAH intensity at the high luminosity end in star-forming galaxies, suggesting that the smallest PAHs may survive in strong radiation fields \citep{Lai2020}. 

SFR can also be estimated from various other indicators such as H$\alpha$. The total SFR estimated from H$\alpha$ using the JWST NIRSpec is $\sim$1700 M$_{\odot}$ yr$^{-1}$ \citep{Ishikawa2024}. The H$\alpha$-based SFR is consistent with the PAH-based SFR given the typical 1-$\sigma$ systematic error of 0.2 dex. \citet{Shipley2016} found a tight correlation between PAH luminosity and the extinction-corrected H$\alpha$ luminosity for 105 galaxies at $0<z<0.4$ over a wide range of luminosities, suggesting that the PAH features may be as accurate an SFR indicator as hydrogen recombination lines. 
%SFR based on far-infrared luminosity for nearby quasars might be systematically biased compared that from H$\alpha$ because of contamination from quasar emission \citep{Husemann2014}. 
The PAH map of J0749+2255 does not show strong contribution from the quasars and no strong quasar outflow is seen, so we do not expect a strong bias of PAH-based SFR due to quasars.

\subsection{Stellar mass - SFR relation and Kennicutt-Schmidt law}

\begin{figure*}
    \centering
    \includegraphics[width=\textwidth]{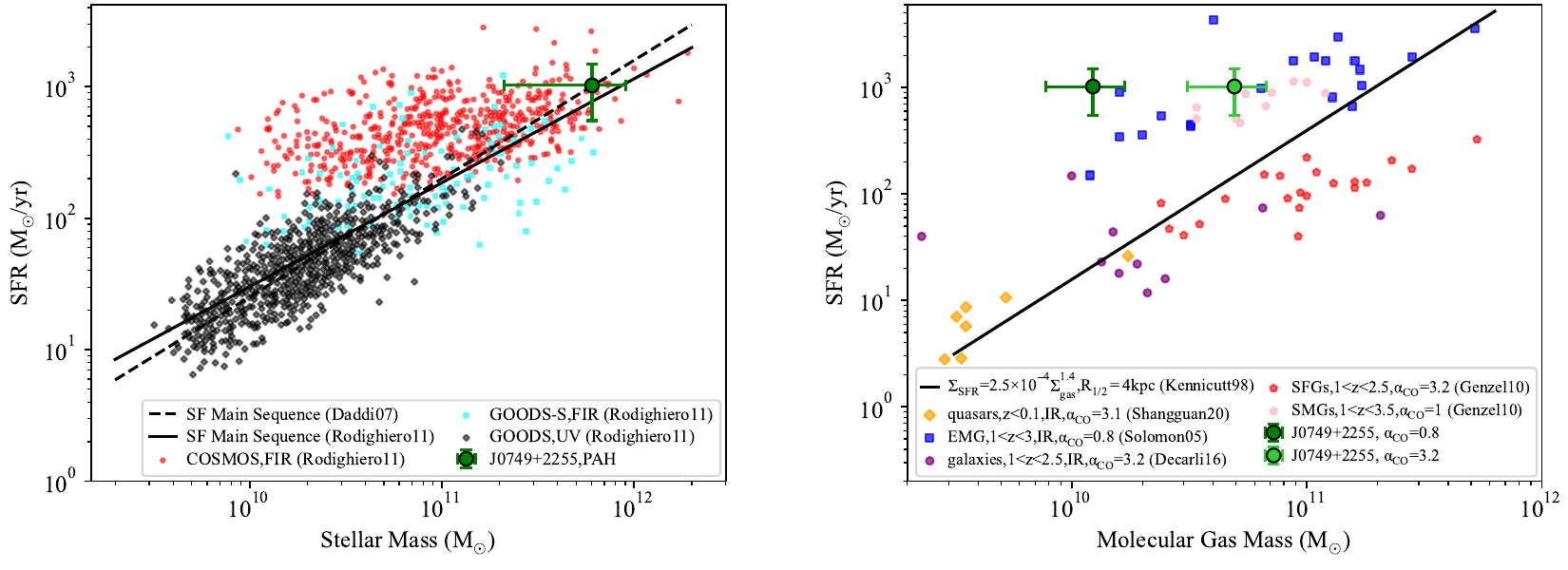}
    \caption{Left: SFR vs. stellar mass of host galaxy for J0749+2255 in comparison to the main sequence of star-forming galaxies. The red circles, cyan squares, and black diamonds are star-forming galaxies at $1.5<z<2.5$ in COSMOS and GOODS fields using UV- or FIR-based SFR \citep{Rodighiero2011}. The black dashed and solid lines represent the main sequence of star-forming galaxies \citep{Daddi2007,Rodighiero2011}. Right: SFR vs. molecular gas (H$_2$+He) mass of the host galaxies for J0749+2255 in comparison to different galaxy/quasars/sub-millimeter galaxy samples. The black line represents the molecular Kennicutt-Schmidt law $\Sigma_{SFR}$=2.5$\times10^{-4}\Sigma_{gas}^{1.4}$ \citep{Kennicutt1998review}. We use the half-light radius $R_{1/2}$=4kpc and calculate the molecular gas mass M$_{mol-gas}$ = 2$\times\pi R_{1/2}^2\Sigma_{mol-gas}$.
    The orange diamonds are local PG quasars \citep{Shangguan2020}. The blue squares are emission line galaxies (EMGs) at $1<z<3$ \citep{Solomon2005}. The red and pink hexagons are star-forming glaxies (SFGs) and sub-millimeter galaxies (SMGs) at $1<z<3.5$ \citep{Genzel2010}. The purple circles are CO-detected galaxies at $1<z<2.5$ in the Hubble Ultra Deep Field \citep{Decarli2016}. The CO-to-H$_2$ conversion factor $\alpha_{CO}$ used in each sample is also shown. }
    \label{fig:SFR-stellar_gas}
\end{figure*}

We can combine the SFR estimated from PAH 3.3\um\ with other physical properties of host galaxies (e.g., stellar mass and molecular gas mass) to investigate whether kpc dual quasars like J0749+2255 have enhanced or suppressed SFR. Using the estimated combined host galaxy mass of 10$^{11.78}$M$_{\odot}$ from the F160W images obtained with the \hst\ \citep{ChenYC2023a}, we derive the specific SFR as 1.7$\times$10$^{-9}$ yr$^{-1}$. In \autoref{fig:SFR-stellar_gas}, left, compared to the empirical stellar mass - SFR relations of single galaxies, J0749+2255 resides on the luminous end of the main sequence for star-forming galaxies at $z\sim2$ \citep{Daddi2007,Rodighiero2011}. The highly star-forming nature of J07492+2255 is very intriguing because our target selection is not based on any of the galaxy properties.

In addition to exploring the stellar mass–SFR relationship, we also investigate whether J0749+2255 adheres to the molecular gas Kennicutt-Schmidt law \citep{Schmidt1959,Kennicutt1998review,Kennicutt2012review}. The molecular hydrogen H$_2$ mass is derived from recent ALMA CO (4-3) observations (Ishikawa et al. in prep.). We obtain the total line-integrated flux $S_{CO}\Delta v$ of J0749+2255 in unit of Jy km s$^{-1}$ and convert it to the intrinsic CO luminosity $L'_{CO}$ using the following equation:
\begin{equation}
    L'_{CO} = 3.25\times10^7 S_{CO}\Delta v \frac{D_L^2}{(1+z)^3\nu_{obs}^2}  \textrm{K kms}^{-1} \textrm{pc}^2,
\end{equation}
where $D_L$ is the luminosity distance in Mpc, $z$ is the redshift, and $\nu_{obs}$ is the observed frequency in GHz.
We assume that the low-J CO transitions are thermalized and optically thick, so $R_{\textrm{41}}=L'_{\textrm{CO 4-3}}/L'_{\textrm{CO 1-0}}\sim1$. We use $R_{\textrm{41}}=0.87$ obtained from local quasars \citep{Carilli2013}. We also include a factor of 1.36 for helium to compute the molecular gas mass.  We assume the CO luminosity-to-H$_2$ mass conversion factor $\alpha_{CO}$ of 0.8 \citep{Solomon2005,Tacconi2008}, though the systematic uncertainty could be at least 30\% \citep{Bolatto2013,Papadopoulos2012}.
%\textcolor{red}{[or typical 3.6 \citep{Daddi2010}]}. 
The molecular gas mass can be calculated as 
\begin{equation}
    M_{mol-gas} = 1.36\alpha_{CO}R_{\textrm{41}}^{-1}L'_{\textrm{CO 4-3}}.
\end{equation}
The estimated molecular gas mass is 10$^{10.09\pm0.16}$ M$_{\odot}$. We plot SFR as function of gas mass for J0749+2255 (\autoref{fig:SFR-stellar_gas}), in comparison to local quasars \citep{Shangguan2020}, various galaxy samples at $z\sim2$ \citep{Solomon2005,Genzel2010,Decarli2016}, and the molecular Kennicutt-Schmidt law \citep{Kennicutt1998review}. 
We find that J0749+2255 exhibits higher ($\sim$ 5-20 times) SFR compared with those samples with the same amount of molecular gas. Even if we consider higher CO-H$_2$ conversion factor $\alpha_{CO}$ of 3.2, J0749+2255 is still an outlier from the Kennicutt-Schmidt law at its redshift. 

We discuss the possible reasons for the deviation. PAH-based SFR could be biased \citep{Spilker2023}, however, the good agreement between PAH-based and H$\alpha$-based SFR makes this scenario unlikely. Even if PAHs are affected by quasars, the feedback from quasars is usually negative \citep{Xie2019,Diamond-Stanic2010} and we do not observe any vacancy or reduction around the quasars in the PAH map. The destruction of PAH will make the discrepancy even worse.  Therefore, this deviation could originate from the enhancement of SFR or depletion of molecular gas due to mergers or quasars. 
Merging galaxies are thought to trigger episodes of extreme star formation \citep{Barnes2004,Saitoh2009}, though some observations suggest the enhancement of SFR due to mergers are moderate, with a factor of 2-3, or not significant \citep{Ellison2013,Kanapen2015,Silva2018,Pearson2019}. Quasars have been known to both suppress and trigger star formation in their host galaxies \citep{Cresci2018}. 
The outflow from quasar could also destroy or expel the molecular gas reservoir \citep{Schawinski2009}. Nevertheless, the outflow from quasars is weak in J0749+2255 based on the \feii\ map and the \oiii\ analysis \citep{Ishikawa2024} , thus the contribution from quasar might not be significant. One potential explanation is that the elevated stage of star formation has persisted for an extended duration, without diminishing even as the molecular gas reservoir is gradually depleted.
%But we do not observe a strong quasar component in the \feii map and , usually though as a shock diagnostic, and the outflow from quasar is weak \citep{Ishikawa2024}. 

%Though we do not observe a strong quasar (PSF) component in the PAH map, it is possible that the enhancements might be extended, rather than only the central region close to quasars. However, J0749+2255 is well aligned with the main sequence of star-forming galaxies in the SFR-stellar mass diagram, suggesting that quasars contribution to SFR might not be huge. Depletion of molecular gas due to quasars is likely a more plausible cause. The high-energy photons from quasars could ionize or dissociate. The NIRSpec IFU study also shows a photoionization region around the northwestern quasar in the BPT diagram \citep{Ishikawa2024}. 

To sum up, the high SFR rate of J0749+2255 suggest possible enhancement of SFR during the dual quasar stage. This starburst activities could be supported by simulations or calculations showing enhanced star formation in interacting galaxies \citep{Kanapen2015,Moreno2019} or during the quasar stage \citep{King2005}. Though whether two quasars are a result of a merger or born in the same massive galaxy is still in debate \citep{Ishikawa2024}, the connection between dual quasars and high SFR is striking because our target selection is not based on galaxy properties.

% https://ned.ipac.caltech.edu/level5/March15/Kennicutt/Kennicutt6.html#Figure%2011

\subsection{[Fe II] is likely driven by star formation} \label{sec:dis:feii}

\feii\ lines observed at the IR wavelengths are commonly associated with shock-heated nebulae, such as supernova remnants  and young stellar objects \citep{Greenhouse1991,Nisini2002}. 
These shocks can destroy grains, causing iron atoms in the dust grain to be sputtered into the gas phase and ionized by the radiation field \citep{Mouri2000}. Quasar-driven outflows may also generate shocks when interacting with the neutral interstellar medium, leading to strong \feii\ emission \citep{Hill2014}. Our aim here is to examine whether the observed \feiifull\ line is attributed to quasars or star formation within the host galaxies. To address this, we utilize the star formation rate derived from the 3.3\um\ PAH to calculate the supernova rate, and subsequently convert it into an anticipated \feii\ luminosity originating solely from the host galaxy star formation.

%We compare the SFR estimated from \feiifull\ with that from PAH 3.3\um, which is solely from the host galaxies, and see if \feii-based SFR show any excess, possible from \citep{Peeters2004}.
%($\Gamma=1.35$ or $\alpha=2.35$) 

The relationship between SFR and SN rate $\nu_{SN}$ in the host galaxy can be expressed by
\begin{equation}
    \frac{\nu_{SN}}{SFR} = \frac{\int^{M_U}_{M_{L,SN}}  \frac{{\rm d}N}{{\rm d}M} {\rm d}M}{\int^{M_U}_{M_{L,SFR}} M\times \frac{{\rm d}N}{{\rm d}M} {\rm d}M},
\end{equation}
where $M$ is the mass of a star, $\frac{{\rm d}N}{{\rm d}M}$ is the initial mass function, $M_{L,SN}$ is the lower mass limit of stars that become SN, and $M_{L,SFR}$ is the lower mass limit of stars that contribute to the SFR. We assume that the PAH-based SFR arises from all stars with a mass higher than 2 $M_\odot$\citep{Peeters2004}, that stars with a mass higher than 8 $M_\odot$ eventually lead to supernovae, and that the initial mass function is truncated at $M_U=100M_{\odot}$.
Using the Salpeter initial mass function with ${\rm d}N/{\rm d}M \propto M^{-\alpha}$ with $\alpha=-2.35$ \citep{Salpeter1955}, the estimated SN rate based on PAH-derived SFR is 27 $yr^{-1}$. 

\citet{Rosenberg2012} find a tight relation between \feii\ 1.26\um\ and SN rate in a sample of nearby galaxies. By incorporating the estimated SN rate into this correlation, we derive the anticipated \feii\ 1.26 \um\ luminosity of $L_{{\rm \feii 1.26}\mu m}$=10$^{8.6}$ L$_{\odot}$, originating solely from star-forming regions within the host galaxies. While the relative strength of the \feii\ 5.34\um\ line and the \feii\ 1.26\um\ lines remains uncertain for J0749+2255, observations of supernova remnants \citep{Reach2006} and simulations of radiative shocks \citep{Hartigan2004} suggest that these lines should have comparable fluxes. We estimate the systematic uncertainty of 0.4 dex for the expected \feiifull\ luminosity using different assumptions. Under the assumption of equal fluxes between the \feii\ 5.34\um\ and the \feii\ 1.26\um\, the observed \feiifull\ luminosity of 10$^{8.7}$ L$_{\odot}$ agrees well with the anticipated \feii\ luminosity of 10$^{8.6}$ L$_{\odot}$ originating from star-forming regions in the host galaxy.
Furthermore, the kinematics of \feii\ presented in \autoref{fig:feii} are consistent with the rotational disk morphology observed in H$\alpha$ \citep{Ishikawa2024}. Hence, we conclude that the \feiifull\ emission of J0749+2255 is primarily associated with star formation rather than with the quasars. Furture JWST observations in NIR and MIR wavelengths with various quasar samples will probe the origin of \feii\ and whether \feii\ can be used as a shock/feedback diagnostic.

\subsection{Spatially resolved \feii/PAH map} \label{sec:dis:feii/pah}

\begin{figure}
    \centering
    \includegraphics[width=0.6\columnwidth]{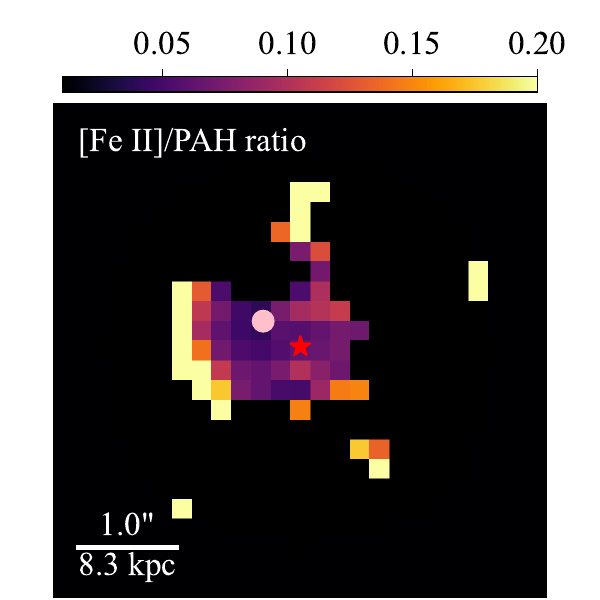}
    \caption{\feiifull/PAH 3.3\um\ flux ratio map. Spaxels close to the edges of IFU are masked to avoid the artifact features. Spaxels with \feii\ signal-to-noise ratio (S/N)  $<$2 are masked. North is up, and east is to the left.}
    \label{fig:Feii_PAH}
\end{figure}

Utilizing the spatially resolved \feiifull\ and PAH 3.3\um\ map of J0749+2255, we aim to explore the spatial variations in the \feii/PAH ratio. To this end, we reproject and convolve the PAH map with a 2D Gaussian kernel to match the pixel scales and the PSF size in the \feii\ map. The resulting \feii/PAH ratio map is shown in \autoref{fig:Feii_PAH}. According to the calculations in Section \ref{sec:dis:feii}, the anticipated \feii/PAH flux ratio from star formation is approximately 0.16. In the host galaxy of J0749+2255, the \feii/PAH ratios range between 0.05 and 0.2, broadly consistent with the expected value. 
%Bright edges ten kpc east and north of J0749+2255 are likely artifacts in the \feii\ map, based on visual inspection of the \feii\ spectra. 
Contrary to expectations, there is no observable increase in \feii/PAH ratios at both nuclei, as would be anticipated from enhanced \feii\ emission because of quasar-driven shocks \citep{Hill2014} or the disruption of small PAH molecules by the strong radiation from quasars \citep{Diamond-Stanic2010,Wu2010}. 
We conclude that neither the radiation field of the quasars, nor quasar-driven outflows penetrate sufficiently into the gas-rich host galaxy and therefore there is no detectable spatially resolved sign of either radiative suppression of PAHs or quasar-driven shocks that would enhance \feii. This is in line with other observations of J0749+2255 which demonstrate the lack of evidence from quasar-driven outflow tracers such as \oiii\ \citep{Ishikawa2024}.

%the lack of evidence from quasar-driven narrow-line-region outflow tracers such as \oiii. The spatially resolved map of \oiii\ from JWST exhibits compact weak signals and small radial velocities of $\lesssim$250 km/s \citep{Ishikawa2024}.

%In Section \ref{sec:dis:feii}, we conclude that most \feii\ are likely from SN in host galaxies. 

\section{Conclusions} \label{sec:conclusion}
In this study, we present high spatial resolution MIR IFU observations of a 3.8-kpc dual quasar at $z=2.17$. Our analysis involves extracting the MIR continuum luminosity for each quasar, leading to the construction of their SEDs. The SEDs of both quasars align with the characteristics of typical optically-selected radio-loud quasars (\autoref{fig:SED}). Additionally, we identify and characterize strong spatially extended features, including the PAH 3.3\um\ feature and the \feiifull\ line, as shown in \autoref{fig:pah} and \autoref{fig:feii}, respectively. We see the rotation of the host galaxy in the \feiifull\ map, consistent with that seen in other tracers (e.g., \ha). 
%The \feiifull\ map reveals distinct blueshifted and redshifted components, with radial velocities of up to $\sim$200 km/s along the NW-SE direction.

By leveraging the PAH 3.3\um\ feature, we estimate a SFR of $10^{3.0\pm0.2}$ M$_{\odot}$ yr$^{-1}$ for J0749+2255. When combined with the stellar mass, the specific SFR of J0749+2255 is 1.7$\times$10$^{-9}$ yr$^{-1}$, placing it in a category similar to star-forming galaxies at redshift $z\sim2$ (\autoref{fig:SFR-stellar_gas}), which is exceptional given J07492+2255 was not selected based on its host galaxy properties. Using the molecular gas mass estimated from CO, we find that the SFR of J0749+2255 is $\gtrsim$10 times higher than the molecular Kennicutt-Schmidt law or the comparison samples with the same amounts of  molecular gas. The deviation could be related to the prolonged stage of high star formation, persisting even as the molecular gas reservoir is depleted. %We think the depletion of molecular gas, because of quasars' photionization, is the most plausible cause. 

To investigate the origin of the \feii\ emission — whether driven by star formation or quasar outflows — we calculate the expected \feii\ luminosity solely from star formation and compare it with the observed value. The observed and expected \feii\ luminosities agree, suggesting that the predominant source of \feii\ emission is likely star formation within the host galaxy. Additionally, no noticeable small-scale quasar-driven wind signatures are observed in the spatial variation of the \feii/PAH map (\autoref{fig:Feii_PAH}). We do not detect any rise in \feii/PAH ratio around quasars that could be related to quasar-driven wind. Based on our analysis of \feii\ and PAH, we conclude that both quasars in J0749+2255 do not exhibit strong outflows that significantly impact the host galaxy, consistent with observations from other outflow tracers. 

In summary, we discover that the 3.8 kpc dual quasar, J0749+2255, resides in a powerful starburst galaxy using the PAH 3.3\um\ observation. 
%In summary, our study demonstrates the capability of \jwst\ to conduct spatially resolved MIR observations for kpc-scale dual quasars, facilitating an exploration of their connections with host galaxies.
The extremely high SFR reveals a possible connection between star formation activities and dual quasar phase. The lower molecular gas mass in J0749+2255, compared with the molecular Kennicutt-Schmidt law, suggests the elevated stage of star formation
might have persisted for an extended duration, even after the molecular gas reservoir is depleted. Our study demonstrates the capability of \jwst\ to conduct spatially resolved MIR observations for kpc-scale dual quasars. We anticipate that future \jwst\ MIR observations on a larger sample of dual quasars will provide a statistically robust understanding of how kpc-scale dual quasars influence star formation and molecular gas within host galaxies.

\begin{acknowledgments}

This work is based on observations made with the NASA/ESA/CSA James Webb Space Telescope. The data were obtained from the Mikulski Archive for Space Telescopes at the Space Telescope Science Institute, which is operated by the Association of Universities for Research in Astronomy, Inc., under NASA contract NAS 5-03127 for \jwst. These observations are associated with program \#2654.
Support for programs GO-02654 and ERS-01335 (NLZ, DR, AV, SV, SS) was provided by NASA through grants from the Space Telescope Science Institute, which is operated by the Association of Universities for Research in Astronomy, Inc., under NASA contract NAS 5-03127.
This work is supported by the Heising-Simons Foundation and Research Corporation for Science Advancement, and NSF grant AST-2108162 (XL, YS, AG).

\end{acknowledgments}

%% To help institutions obtain information on the effectiveness of their 
%% telescopes the AAS Journals has created a group of keywords for telescope 
%% facilities.
%
%% Following the acknowledgments section, use the following syntax and the
%% \facility{} or \facilities{} macros to list the keywords of facilities used 
%% in the research for the paper.  Each keyword is check against the master 
%% list during copy editing.  Individual instruments can be provided in 
%% parentheses, after the keyword, but they are not verified.

\vspace{5mm}
\facilities{JWST(MIRI)}

%% Similar to \facility{}, there is the optional \software command to allow 
%% authors a place to specify which programs were used during the creation of 
%% the manuscript. Authors should list each code and include either a
%% citation or url to the code inside ()s when available.

\software{astropy \citep{Astropy2013,Astropy2018,Astropy2022}, reproject \citep{reproject2023}
\texttt{q3dfit} \citep{Rupke2014-ifsfit,Rupke2021-questfit}          }

%% Appendix material should be preceded with a single \appendix command.
%% There should be a \section command for each appendix. Mark appendix
%% subsections with the same markup you use in the main body of the paper.

%% Each Appendix (indicated with \section) will be lettered A, B, C, etc.
%% The equation counter will reset when it encounters the \appendix
%% command and will number appendix equations (A1), (A2), etc. The
%% Figure and Table counter will not reset.

\appendix

%% For this sample we use BibTeX plus aasjournals.bst to generate the
%% the bibliography. The sample631.bib file was populated from ADS. To
%% get the citations to show in the compiled file do the following:
%%
%% pdflatex sample631.tex
%% bibtext sample631
%% pdflatex sample631.tex
%% pdflatex sample631.tex

\bibliography{ref_new}{}
\bibliographystyle{aasjournal}

%% This command is needed to show the entire author+affiliation list when
%% the collaboration and author truncation commands are used.  It has to
%% go at the end of the manuscript.
%\allauthors

%% Include this line if you are using the \added, \replaced, \deleted
%% commands to see a summary list of all changes at the end of the article.
%\listofchanges

\end{document}